# Template-Free Preparation of Thermoresponsive Magnetic Cilia Compatible with Biological Conditions

Aline Grein-Iankovski,[†,‡] Alain Graillot,[¶] Milad Radiom,[†] Watson Loh,*[,‡] and Jean-François Berret*[,†]

[†]Université de Paris, CNRS, Matière et systèmes complexes, 75013 Paris, France.
[‡]Institute of Chemistry, University of Campinas (UNICAMP), P.O. Box 6154, 13083-970 Campinas, Brazil.
[¶]Specific Polymers, ZAC Via Domitia, 150 Avenue des Cocardières, 34160, Castries, France
E-mail: wloh@unicamp.br; jean-francois.berret@univ-paris-diderot.fr

**Abstract**
Bio-inspired materials are commonly used in the development of functional devices. The fabrication of artificial cilia mimicking the biological functions has emerged as a promising strategy for fluid manipulation in miniaturized systems. In this study, we propose a different physicochemical insight for the preparation of magnetic cilia based on the temperature-triggered reversible assembly of coated iron oxide nanoparticles in a biocompatible template-free approach. The length of the prepared cilia could be tuned between 10 – 100 µm reaching aspect ratios up to 100 in a very dense array of flexible structures with persistence lengths around 8 µm. Magnetic actuation of the cilia revealed robust structures (over several hours of actuation) with a wide range of bending amplitudes resulting from high susceptibility of the filaments. The results demonstrate that the proposed strategy is an efficient and versatile alternative for templated fabrication methods and producing cilia with remarkable characteristics and dimensions within the template-free approaches.

# Introduction

Nature has plenty of complex processes and remarkable structures whose performances in- spire the development of new materials and innovative functional devices. Cilia and flagella are examples of essential structures that provide a range of vital functions in varied biological organisms.[1] Generally, these structures present high aspect ratios (ratio of the length to the diameter) with diameter around 0.15 – 0.30 µm and lengths varying from few micrometers up to millimeter range.[1] In bacteria, these surface-attached organelles are intrinsically related to motility in different ways, including propulsion, self-locomotion and fluid transportation inside the human body.[2] In the bronchial region of the lungs, the collective beating motion and the hydrodynamic coupling of cilia through the cell membrane generates a wave-like pattern, called metachronal waves, whose beating moves mucus towards the pharynx and expel it from the airways.[1,3]

Inspired by the efficiency of these biological cilia, a variety of artificial structures have been proposed to mimic their behavior in many applications, for example, for fluids manipulation in minia-turized devices, in microfluidic and lab-on-a-chip devices.[4,5] Among the strategies explored to induce motion in these ciliary arrays, we emphasize the magnetic actuation due to their rapid response and efficient non-contact and non-intrusive control over the system. Several theoretical approaches have already investigated the mechanism of the ciliary synchronized motion and demonstrated the possibility of artificial cilia to effectively generate a directional flow for controlling the motion in fluid environments.[6,7] Notable experimental efforts have been made in this direction to prepare biomimetic magnetic cilia using different types of magnetic particles embedded in a soft matrix by different fabrication approaches. A range of nicely shaped cilia arrays with controlled magnetic actuation have been fabricated, many of them displaying great potential for fluid and particle manipulation.[8–30] However, most of these studies are based on patterning, lithography or rather tedious fabrication processes, even requiring additional steps for the mold preparation. Also, the geometry of the cilia is limited to the pattern of the available mold and to its fabrication boundaries, especially regarding the size, restricting the adjustment of the structures in the array. Therefore, the current limitations regarding cilia size, aspect ratio and density, combined with the desirable search for versatile preparation procedures offer opportunities to explore and contribute to improvements in this field.

In this work, we propose a novel insight within the template-free strategies for the preparation of magnetic cilia exploring the thermoresponsive phase transition of polymer decorated magnetic nanoparticles to induce an oriented assembly. When an external magnetic field is applied over a colloidal suspension of magnetic particles it orients the individual magnetic moment carried by the nanoparticles inducing dipole-dipole interparticle attraction, leading to chain formation.[31–36] By orienting this assembly perpendicular to a substrate, connecting the particles and anchoring the formed chains on the surface it is possible to produce a magnetic ciliary array. So far, a couple of template-free approaches have been proposed base on the alignment of magnetic beads,[8,14,16] carbonyl iron particles,[26,28,37] and cobalt particles,[11,21,25] or nanoparticles,[13] for example. The explored strategies for linking these particles varied from covalent binding via amidation chemistry,[8,13] electrostatic attraction,[14,16,38] dipolar assembly,[13] and incorporation in an elastomeric matrix.[11,21,22,25,26,28,37]. The moldless state of magnetic fabrication enables the preparation of cilia with much higher aspect ratios and the rough adjustment of their dimensions and density by modulating the magnetic field, by varying the particles concentration or the particles size.[11,13,14,21,22,39] Cilia with lengths up to 6000 μm with an average aspect ratio of 120 were obtained,[11] although, cilia with sub- micrometer diameters, within the size range of biological cilia, are only achievable employing nanometer size particles.[13]

Our contribution accomplishes the temperature-triggered fabrication of high aspect ratio cilia composed of polymer-coated iron oxide nanoparticles. The thermoresponsive polymer at the nanoparticle surface acts as a linker, promoting upon heating nanoparticle binding into elongated filaments. The transition occurs at biological compatible temperature (around 37 °C) and ionic strength (0.1 – 0.2 M) and within a broad pH range (2 – 8). The resulting magnetic cilia present remarkable flexibility with persistence lengths around 8 µm and aspect ratios up to 100 with submicrometer diameters. Their lengths can vary from 10 – 100 µm by increasing the concentration of the nanoparticles, which also results in a higher coverage density (up to 0.1 filament/µm$^2$).

# Experimental section

## Materials

The superparamagnetic maghemite nanoparticles ($\gamma$-Fe$_2$O$_3$) with mean diameters of 8 nm and 14 nm were previously synthesized by the coprecipitation method.[40] The statistical propylacrylamide copolymer bearing phosphonic acid groups (PAmPh) was custom synthesized by Specific Polymers (Montpellier, France). The synthesis was performed by free radical polymerization using AIBN as radical initiator and the details can be found in the work of Graillot *et al.*[41] The copolymer is composed by thermosensitive N-n-propylacrylamide monomers (NnPAAm [25999-13-7] – SP43-0-002) and di-methoxyphosphoryl)methyl 2-methylacrylate (MAPC1, [86242-61-7] – SP41-003) monomers in the molar ratio (0.98:0.02), respectively. The copolymer chemical structure is presented in the Fig. S1 (Supporting In- formation). The number-averaged molecular weight of the copolymer is $M_n$ = 49500 g mol$^{-1}$ with a dispersity of Đ= 2.8, rendering an average of 8 phosphonic acid groups per polymer chain. Phosphate buffered saline (PBS), nitric acid (HNO$_3$), ammonium hydroxide (NH$_4$OH) and Pluronic F-127 were purchased from Sigma-Aldrich. Purified water was obtained with a Milli-Q purification system.

## Preparation of thermoresponsive magnetic nanoparticles

The ($\gamma$-Fe$_2$O$_3$) nanoparticles were coated exploring the binding affinity of phosphonic acid groups to metal oxide surfaces.[42] Initially, 0.1 *wt%* solutions of both the copolymer PAmPh and $\gamma$-Fe$_2$O$_3$ nanoparticles were prepared at pH 2.0 (adjusted with HNO$_3$) and filtered with a 0.22 *µ*m pore membrane (Millipore). All the preparation procedure was temperature controlled under 10 °C to avoid the thermal transition and aggregation of the copolymer. The $\gamma$-Fe$_2$O$_3$ solution was added dropwise to the copolymer solution under vigorous stirring until reach a mixing volume ratio of 0.5, ensuring enough polymer chains to well cover all the nanoparticles surface. In the sequence, the pH was raised to 7.5 with the addition of NH$_4$OH and the hybrid $\gamma$-Fe$_2$O$_3$@PAmPh nanoparticles were dialyzed in 100 kDa tubing membranes (Spectra/Por™) for four days at 4 °C to remove the excess of non-linked copolymers. The dialyzed nanoparticles were finally centrifuged at 4000 rpm in 100 kDa centrifuging filters (Merck) to reach a final concentration of 0.1 *wt%* $\gamma$-Fe$_2$O$_3$@PAmPh. The colloidal stability of the $\gamma$-Fe$_2$O$_3$@PAmPh and their thermal transition were characterized by dynamic light scattering and zeta potential measurements using a Malvern NanoZS Zetasizer instrument.

## Fabrication of magnetic cilia

For the preparation of magnetic cilia, 0.005 *wt%*, 0.01 *wt%* and 0.05 *wt%* $\gamma$-Fe$_2$O$_3$@PAmPh dispersions were prepared in cold phosphate buffered saline (PBS, pH 7.4, 0.01 M phosphate buffer, 0.0027

M potassium chloride and 0.137 M sodium chloride) and kept refrigerated under 10 °C until usage. 25 μL of the γ-$Fe_2O_3$@PAmPh solution were deposited on glass slides containing an adhesive Gene Frame (Abgene) of 1 cm x 1 cm dimensions and 250 μm thickness, and then sealed with a cover slip. The prepared glass slide containing the sample was placed on the center of a 10 cm x 10 cm x 1 cm Ferrite magnet and an identical magnet was placed over it keeping a gap of 0.6 cm. The resulting field is 40 mT in the center of the magnet. The setup was heated to 50 °C for 30 minutes, cooled down to room temperature before removing the sample from the magnetic field. It is noteworthy that the formation of the elongated structures is already observed at temperatures around 37 °C in less than 10 minutes. Higher temperatures and longer times favor a stronger assembly between the nanoparticles, prolonging the cilia lifetime and facilitating their characterization. A schematic representation of the experimental setup and assembly steps for the preparation of cilia is presented in Fig. 1.

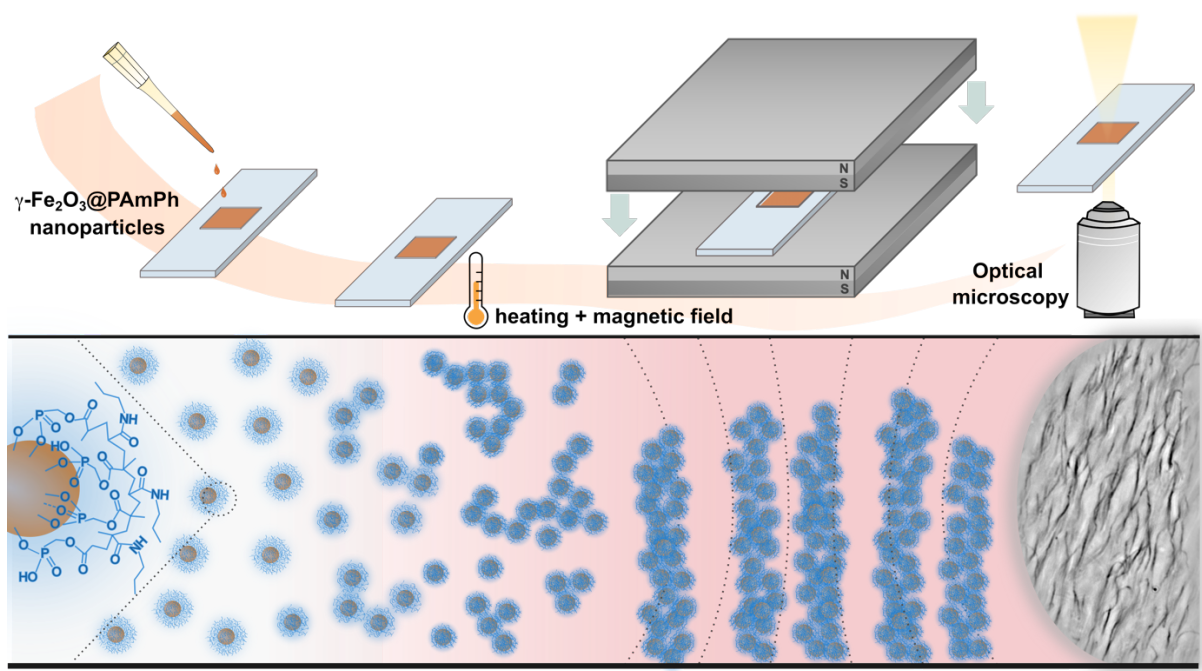

***Figure 1:*** *Scheme of the experimental steps used for the cilia preparation using thermoresponsive iron oxide nanoparticles γ-$Fe_2O_3$@PAmPh (top) and a microscopic illustrative representation of the transition the system undergoes upon heating in the presence of a magnetic field (bottom).*

### Characterization

The prepared cilia were analyzed using an IX73 Olympus inverted microscope equipped with 40× objective lens coupled with a CCD camera (QImaging, EXi Blue). The images and videos were captured in the phase contrast mode supported by the software Metaview (Universal Imaging) and analyzed using the ImageJ software. The magnetic actuation of the cilia was induced by controlling the position of a permanent Neodymium magnet (0.4 T at the magnet surface) around the sample (3 – 6 cm apart). The transmission electron microscopy (TEM) images were obtained on a Jeol-100 CX microscope. The nanoparticle and filament dispersions were deposited on 400 mesh copper grids (Ted Pella) and air dried. Scanning electron microscopy (SEM) analysis were performed at 5K kV using a Zeiss Supra 40 MEB-FEG instrument. For the calculation of cilia persistence lengths ($L_P$) we imaged several isolated filaments prepared at c = 0.01 *wt*% and freely adsorbed on glass surfaces. The data was analyzed with MATLAB software (MathWorks Inc.), where the contour of the filaments were traced and the $L_P$ values were calculated using the tangent vector correlation function (Eq. 1).

$$\langle \cos \theta_{s,s+L} \rangle = exp\left(\frac{-L}{2L_P}\right) \qquad (1)$$

where $\vartheta$ is the angle between tangent vectors at points s and s+L, by averaging s along the contour.[43,44] The cilia robustness against an imposed flux was investigated in a flow cell, prepared by adapting an inlet and outlet channels in both sides of the framed glass slide. The cilia were prepared in the closed cell and afterwards the inlet channel connected to a syringe pump (New Era, NE-1000). A controlled water flux was applied at different rates and the robustness of the cilia analyzed by optical microscopy.

The ability of the magnetic cilia to remotely induce motion of a fluid was investigated. In one experiment an aqueous dispersion of 5 $\mu$m silica particles (Kromasil, 100 Å pore size) was injected (flux of 10 $\mu L\ min^{-1}$) in the cell containing a fresh prepared cilia (0.005 wt%). Similarly, in a second approach 25 $\mu$m emulsion droplets of silicon oil in water (stabilized by Pluronic F-127) were injected over a cilia array (0.005 wt%). Afterwards, the cilia arrays containing the solid particles and oil droplets were actuated using a permanent Neodymium magnet (0.4 T at the magnet surface) in horizontal oscillating movements over the sample (3 – 6 cm apart) and simultaneously analyzed in the optical microscope as specified before.

## Results and discussion

The thermoresponsive magnetic nanoparticles were prepared using a multiphosphonic acid coating strategy previously established in our group.[45] This method is based on the affinity of the phosphonic acid groups to readily bind to the surface of the iron oxide nanoparticles directing the interaction to form a polymer shell.[46] The presence of multiple binding sites along the polymer chain confer a resilient coating and long-term stability to the modified particles.[47] The prepared $\gamma$-Fe$_2$O$_3$@PAmPh presented an average hydrodynamic diameter of 100 nm (PDI = 0.12) and 110 nm (PDI = 0.13) for the samples prepared with 8 nm and 14 nm $\gamma$-Fe$_2$O$_3$ cores, respectively (Fig. S2 - Supporting Information). They displayed a negative surface charge at pH values above 2, indicating the presence of non-linked phosphonic acid groups that deprotonate at higher pH values (pKa$_1$ = 2.13, pKa$_2$ = 7.06).[48] At pH 7.4 the zeta potential of γ-Fe$_2$O$_3$@PAmPh aqueous solution is -30 ± 2 mV and when the nanoparticles are dispersed in PBS solution this value drops to -7 ± 1 mV. The addition of salts minimizes the electrostatic repulsion between the nanoparticles, strengthening their assembly upon heating. Besides, the temperature-induced assembly of γ-Fe$_2$O$_3$@PAmPh can be performed in a broad pH range, from 2 – 8 in saline media (0.1 – 0.2 M).

The propylacrylamide monomers confer the thermoresponsive properties to the copolymer.[41] The characteristic low critical solution temperature (LCST) transition of this copolymer when immobilized at the γ-Fe2O3 nanoparticles surface is around 25 °C in PBS (Fig. S3 – Supporting Information). Below LCST, the formation of a structured hydration layer around the polymer chains stabilizes the nanoparticles, conferring colloidal stability to the γ-Fe2O3@PAmPh dispersion at low temperature. As the temperature increases above its LCST, this surrounding water network is no longer stable and the hydrophobic inter- actions between the polymer monomers become dominant causing a coil-to-globule phase transition.[49] The collapse of the polymer shell leads to the nanoparticle aggregation and the attachment of the formed filaments to the surface. Their adhesion to glass is promoted by physical adsorption and reinforced by hydrogen bonding and hydrophobic interaction as the polymer collapse upon heating. In addition, the magnetic attraction force pulls the filaments towards the hot surface strengthening their interaction.

Fig. 2 shows electron microscopy images of the system at different stages of the assembly process towards the formation of elongated filaments. Initially, the γ-Fe$_2$O$_3$@PAmPh nanoparticles are well dispersed below the polymer LCST with an average core diameter of 8 nm (Fig. 2a). The observed nanoparticle clusters are a drying artifact. As the system is heated to 37 °C, above the LCST, the higher contrast of the collapsed polymer around the nanoparticles is clearly observed, connecting them into larger aggregates (Fig. 2b). When the heating process is performed concomitant with the application of magnetic field, the preformed clusters dis- playing larger magnetic moments assemble together along the field direction growing into elongated filaments (Fig. 2c). These filaments may present some voids and irregularities along their structure, which probably contribute to their flexibility (Fig. 2d). It is note- worthy that it is possible to form the filaments with both 8 nm or 14 nm iron oxide cores and the ciliated surfaces likewise. However, the cilia results shown later in optical microscopy were prepared using 14 nm nanoparticles because they displayed greater contrast, facilitating their visualization.

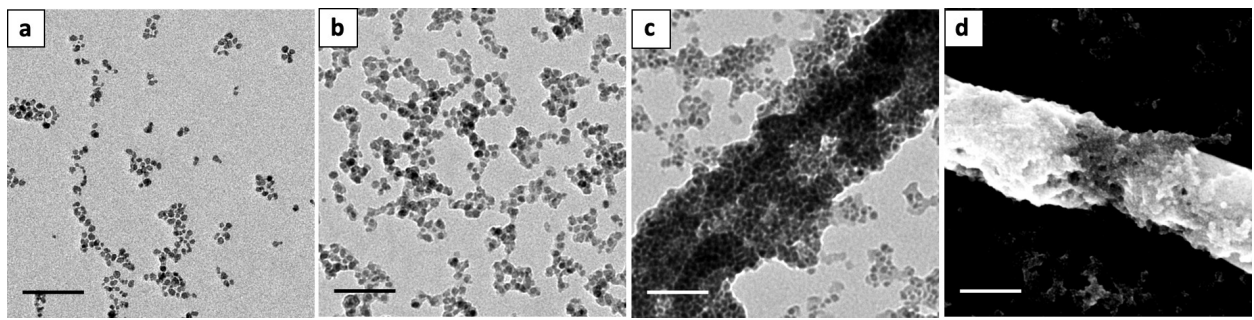

*Figure 2: Steps of the assembly process towards the formation of filaments. TEM images of (a) magnetic γ-Fe$_2$O$_3$@PAmPh dispersion below LCST, (b) γ-Fe$_2$O$_3$@PAmPhh heated to 37°C, (c) γ-Fe$_2$O$_3$@PAmPh aligned and assembled in a magnetic field during heating, Scale bars 100 nm. (d) SEM image of a filament exhibiting local voids in the structure, Scale bar300 nm.*

By changing the concentration of the initial nanoparticle dispersion in the same preparation setup, it is possible to modulate the size of cilia and their coverage density. This is a versatile approach that enables the adjustment of the cilia features according to the required application, to the length scale of the objects to manipulate, to the size of the compartment where they will be inserted, etc. Fig. 3a shows optical microscopy images of the cilia obtained from 0.005 *wt%*, 0.01 *wt%* and 0.05 *wt%* γ-Fe$_2$O$_3$@PAmPh. These images clearly show a sequential growth of the filaments as the nanoparticle concentration increases, as well as the filament crowding on the surface. The insets in the pictures are a top view of the filaments in upright position, showing the difference in the number of attaching spots resulting from a higher coverage density at higher concentrations. We estimate an average increase on the surface coverage density from 0.02 to 0.04 filament/µm$^2$ when the nanoparticle concentration increased from 0.005 to 0.01 *wt%* and reached up to 0.1 filament/µm$^2$ at the highest concentration 0.05 *wt%*.

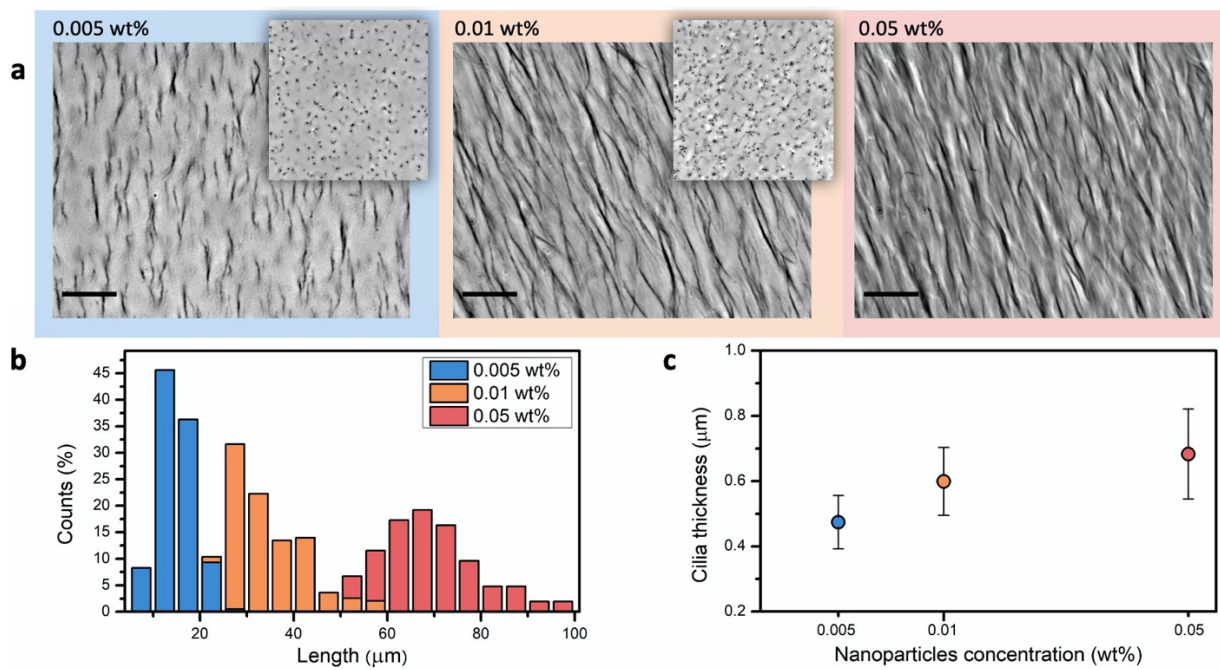

*Figure 3:* Effect of nanoparticles concentration on the dimensions of the formed cilia. (a) Optical microscopy images of the cilia prepared from 0.005 wt%, 0.01 wt% and 0.05 wt% γ-Fe$_2$O$_3$@PAmPh. Scale bars 20 μm. The insets show a top view of the filaments in upright position. Measured cilia size in each case: (b) Cilia length distribution, (c) Average cilia thickness (diameter).

Estimates of the cilia lengths and thicknesses were calculated by measuring 100 – 200 individual cilium and these values are presented in Figs. 3b and 3c, respectively. These estimates show an average increase of 130 % and 350 % in the length and 26 % and 45 % in the thickness of the produced filaments when their concentration increased by 2x and 10x, respectively. The average length of 0.05 *wt*% might be slightly underestimated due to difficulties to focus the entire filaments given their length and crowding. Therefore, these results demonstrate the possibility to obtain cilia with very high aspect ratios, up to 100 in these conditions. The achievable density and aspect ratio ranges obtained in our work are very significant comparing to other magnetic cilia structures reported so far in the literature. Most of the coverage densities obtained for magnetic cilia were below 0.02 filament/μm$^2$.[8,9,11,14] In contrast, Luo et al.[30] obtained a density of 0.25 filament/μm$^2$ for templated cilia with aspect ratios up to 11 and Benkoski et al.[13] prepared 10 μm long cilia with a remarkable aspect ratio of 435 and densities up to 1 filament/μm$^2$. In the present work, we were able to arrange up to 0.1 filament/μm$^2$, with lengths between 10 – 100 μm and aspect ratios up to 100, which are among the highest obtained for magnetic cilia within this lenghtscale.[8,9,13,30] High aspect ratios and high coverage densities usually improve the performance and amplitude of the cilia actuation, making them to behave as a continuous array with high compliance.[30]

These cilia quickly react to an external magnetic stimulus. Their motion can be easily driven ac-cording to the magnetic field direction as can be observed in the Movies 1, 2 and 3 (provided as Supplementary Material) and summarized in Fig. 4. In this case, a permanent neodymium magnet (0.4 T at the magnet surface) was approached to the samples with a gentle oscillating movement, which the cilia promptly followed. As soon as the external field is removed the system relaxes, the cilia structures soften, slowly bend and settle randomly as shown in Movie 4.

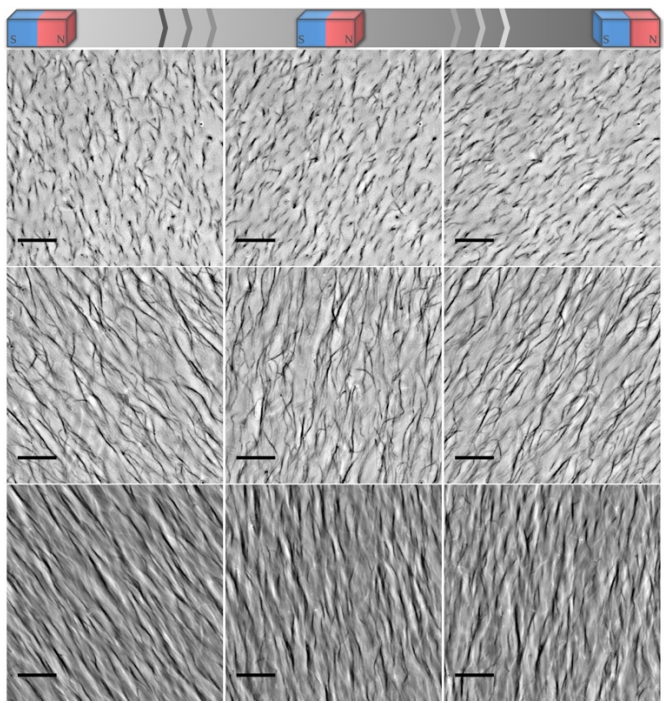

*Figure 4*: *Sequential frames taken from the movies 1, 2 and 3 showing the motion of the cilia in response to the movement of an external magnet. From top to bottom: cilia obtained from 0.005 wt%, 0.01 wt% and 0.05 wt% γ-Fe$_2$O$_3$@PAmPh, respectively. Scale bars 20 μm.*

To estimate the mechanical properties of the cilia we determined their persistence length ($L_P$), which is a characteristic length scale related to the bending rigidity of the material.[50] Fig. 5 shows a representative analysis of an isolated filament (Fig. 5a) with the experimental data and fitted tangent correlation function according to Equation 1 (Fig. 5b). The inverse of the exponential decay results in the $L_P$ value. From tracing 36 individual filaments we estimate an average $L_P$ of 7.6 μm ± 2.9 (Fig. 5c), within the same range of values observed in biological actin filaments, for example.[51] The flexibility of our magnetic cilia originates from the polymeric junctions in between the nanoparticles. Besides, the assembly driving force via hydrophobic attraction also plays a role. In contrast, magnetic wires formed by electrostatic induced assembly presented much stiffer structures with $L_P$ values up to 1 m.[52] Analyzing the filaments tip displacement relative to their base in rotating movements (Movie 5 and Movie 6) clearly demonstrates their wide range of bending amplitudes in all directions.

Flexibility and high bending amplitude are important for the purpose of mimicking the biological structures due to their intrinsic capability of generating metachronal waves. In our case, flexible structures were also observed in the denser sample with high aspect ratio cilia, as emphasized in the Movie 7.

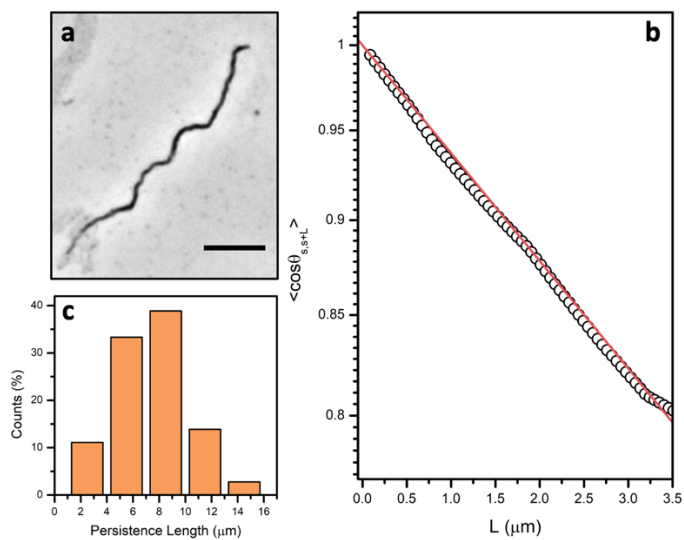

*Figure 5: Analysis of the persistence length of the magnetic filaments. (a) Optical microscopy image of an isolated filament adsorbed on glass used to trace its contour length. Scale bar 10 μm. (b) A representative plot of cos $\vartheta_{s,s+L}$ as a function of the separation L along the contour and the respective exponential fitting. (c) Persistence length distribution obtained from the fitting of individual magnetic filaments.*

The temperature induced assembly of $\gamma$-Fe$_2$O$_3$@PAmPh nanoparticles is a reversible process, although it presents a significant hysteresis on heating/cooling cycles depending on the magnitude of the stimulus employed, with both temperature and time. These cilia, when assembled at 50 °C for 30 minutes, remain functional for a couple of hours after the system is cooled to room temperature. Any movement restriction related to increasing rigidity that could signal flexural fatigue was observed after several cycles of excitation. However, we observed a significant disintegration of the filaments after about 24 hours. Still, their lifetime can be prolonged if the system is maintained above its LCST after preparation. When kept in this condition the cilia remained active for at least 24 hours.

Despite their flexibility, this system can stand the flow of liquids. It is possible to pump water or aqueous solutions through the cilia, for instance, a flux of 10 μL min$^{-1}$. We observed that higher fluxes can cause the entanglement of the cilia and tipping over due to their high aspect ratio. Cilia with a U-shaped geometry are observed in some cases, when the filaments bend and adhere to the surface from their tip extremity. The cilia display leaching out resistance against a flux of 100 μL/min$^{-1}$ for 30 minutes, although significant entanglements and surface adherence were observed in this case. Even though these systems are not permanent, their versatile preparation method compatible with biological conditions is a remarkable advantage, enabling one to envisage the cilia formation in different devices, potentially among living cells and the scaling up of the preparation to larger coverage areas. Also, preliminary experimental evidences confirmed the formation of similar cilia over a metallic film of gold, indicating the possibility to explore different substrates. Using this method, by simply injecting the initial nanoparticles dispersion inside the compartment and locally applying a homogeneous magnetic field and heating, the cilia are promptly formed in a few minutes and are ready for actuation.

In addition, the cilia array was tested for its capacity to impart motion. To demonstrate this, in one experiment a dispersion of 5 μm silica particles and in another experiment an emulsion of 25 μm silicon oil droplets in water were injected over the cilia array formed by 0.05 wt% $\gamma$-$Fe_2O_3$@PAmPh. The cilia arrays were actuated by an oscillating magnetic field and the imparted motions of the solid particles and oil droplets were monitored by optical microscopy. Exemplary Movie 8 shows that cilia move a silica particle in their direction of motion. Similarly, Movie 9 shows the same effect for a silicon oil droplet. However, for an effective displacement, one should optimize the conditions to avoid adhesion to the array.

## Conclusions

In summary, we propose an original approach for the preparation of magnetic cilia based on the temperature induced assembly of iron oxide nanoparticles at biological compatible conditions. This is a versatile template-free methodology that offers the possibility to easily tune the size and coverage density of the cilia by varying the concentration of the nanoparticle dispersion. Structures with lengths between 10 – 100 μm and aspect ratios up to 100 were obtained in a dense array of around 0.1 filament/μm$^2$, among the highest parameters regarding currently reported similar systems. The cilia are obtained in a few minutes and are reproducible and rather robust towards changes in environmental conditions and flux. The structures display high flexibility with persistence lengths around 8 μm and large bending amplitude when actuated by a magnetic field, being able to potentially induce particle motion. This simple protocol could be used over other substrates such as metallic gold, providing a versatile methodology for preparation of responsive magnetic cilia for varied purposes

## Acknowledgement

The authors thank the financial support provided by the São Paulo Research Foundation (FAPESP, grant numbers 2015-25406-5, 2017/04571-3 and 2018/16330-3), ANR (Agence Nationale de la Recherche) and CGI (Commissariat à l'Investissement d'Avenir) through Labex SEAM (Science and Engineering for Advanced Materials and devices) ANR 11 LABX 086, ANR 11 IDEX 05 02. We acknowledge the ImagoSeine facility (Jacques Monod Institute, Paris, France), and the France BioImaging infra-structure supported by the French National Research Agency (ANR-10-INSB-04, « Investments for the future »). This research was supported in part by the Agence Nationale de la Recherche under the contract ANR-13- BS08-0015 (PANORAMA), ANR-12-CHEX-0011 (PULMONANO), ANR-15-CE18-0024-01 (ICONS), ANR-17-CE09-0017 (AlveolusMimics). We gratefully acknowledge the company Specific Polymers (Montpellier, France) in the persons of Cedric Loubat and Nicolas Bia for the providing polymers and for their support. The authors also thank the experimental support from Jerôme Fresnais and the laboratory PHENIX (Université Pierre et Marie Curie, CNRS UMR 8234), Sarra Gam Derouich and the ITODYS SEM facility (Université de Paris, CNRS UMR 7086) and Dyemi Torikai for experimental assistance. We thank the access to microscopy equipment and staff assistance provided by the National Institute of Science and Technology on Photonics Applied to Cell Biology (IN-FABIC) at UNICAMP.## Supporting Information Available

The following files are available free of charge.
- Supporting information: Additional characterization results and detailed captions of the movie files.
- Movie 1: c = 0.005 wt%
- Movie 2: c = 0.01 wt%
- Movie 3: c = 0.05 wt%
- Movie 4: no magnetic field, c = 0.01 wt%
- Movie 5: rotation, c = 0.005 wt%
- Movie 6: rotation, c = 0.01 wt%
- Movie 7: flexibility, c = 0.05 wt%
- Movie 8: particle motion
- Movie 9: oil droplet motion

## References

(1) Sleigh, M. A. In *The biology of cilia and flagella*; Kerkut, G. A., Ed.; Pergamon Press: Oxford, 1962; Vol. 12.
(2) Khan, S.; Scholey, J. M. Assembly, Functions and Evolution of Archaella, Flagella and Cilia. *Current Biology* **2018**, *28*, R278–R292.
(3) Gilpin, W.; Bull, M. S.; Prakash, M. The multiscale physics of cilia and flagella. *Nature Reviews Physics* **2020**, *2*, 74–88.
(4) Fahrni, F.; Prins, M. W. J.; van IJzendoorn, L. J. Micro-fluidic actuation using magnetic artificial cilia. *Lab on a Chip* **2009**, *9*, 3413–3421.
(5) Zhou, Y.; Huang, S.; Tian, X. Magnetoresponsive Surfaces for Manipulation of Non- magnetic Liquids: Design and Applications. *Advanced Functional Materials* **2020**, *30*, 1906507.
(6) Gauger, E. M.; Downton, M. T.; Stark, H. Fluid transport at low Reynolds number with magnetically actuated artificial cilia. *The European Physical Journal E* **2009**, *28*, 231–242.
(7) Elgeti, J.; Gompper, G. Emergence of metachronal waves in cilia arrays. *Proceedings of the National Academy of Sciences* **2013**, *110*, 4470 LP – 4475.
(8) Singh, H.; Laibinis, P. E.; Hatton, T. A. Synthesis of Flexible Magnetic Nanowires of Permanently Linked12

## Graphical TOC Entry

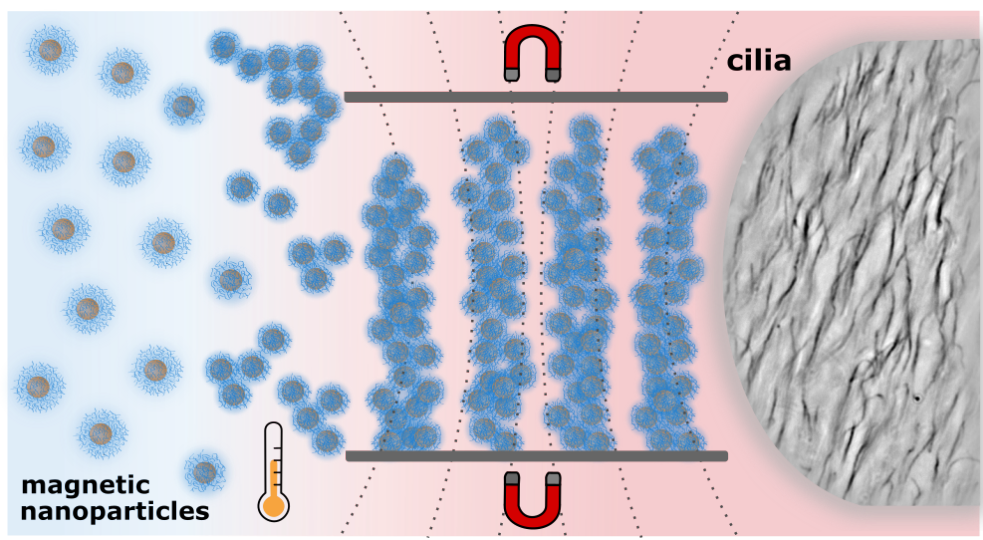